# Shouldn't we make biochemistry an exact science?

August 31, 2014

**Bob Eisenberg**
Department of Molecular Biophysics and Physiology
Rush University Medical Center
Chicago IL
60612

Email Address: *beisenbe@rush.edu*

Exact science is useful. The physics of X-rays is exact. Biochemists can trust X-ray crystallography because the equations of X-rays are exact. But we rarely trust the equations that describe our own experiments and that is for good reason. The equations fail so often.

Biochemists know that the law of mass action we use every day is not exact. The rate constants of that law change as conditions change. When we try to use that law, we must change parameters, but we do not know how. The law of mass action is not exact and not very useful because we cannot transfer it—parameters unchanged—from one set of conditions to another. This fact is known to every enzymologist who measures rate constants, but sad to say, other scientists often are not aware of this reality.

**High resolution calculations do not guarantee useful results**

Biochemists have tried to make their theories exact by increasing resolution. Our models of enzymes include thousands of atoms in cathedrals of structure. The hope has been that computing all the atoms of those cathedrals would produce exact simulations, if not exact equations. But, as the calculations of molecular dynamics reach from atomic to biological scales, we face disappointment once again. Issues of scale in time, space, and concentration must be dealt with all at once, in one calibrated calculation [13]. Enormous resolution does not guarantee useful biological results.[13, 46, 53]

We know very well that most enzyme reactions are controlled biologically by trace concentrations, $10^{-8}$ to $10^{-6}$ M, of ions like $Ca^{2+}$. All atom simulations are not large enough, however, to deal with the 55 M water that dissolves each calcium ion in a $10^{-8}$ to $10^{-6}$ M solution. The atomic resolution of simulations will have limited use if we cannot deal with the trace concentrations that control enzymes in health and disease.

**Force fields of molecular dynamics: boundary charges**

Molecular dynamics almost always uses force fields that depend on the coordinates of only two atoms at a time, calibrated at infinite dilution, i.e., in distilled water. But no two body force field can deal with electric charges that depend on the location of all charges in the system, like polarization charges at the boundaries of systems or at interfaces within the system. The electric force field is defined by the Poisson equation. Dependence on the location of all charges, including boundary charges, is displayed **explicitly** in the general solution of Poisson's equation, see Jackson [32], Section 1.10, specifically eq. (1.36) and eq. (1.42). Boundary charges cannot be neglected in biological problems because they include polarization charges near lipid membranes. Charge at lipid membranes defines the electrical potential of cells and is responsible for the electrical function of nerve and muscle, and all other cells, for that matter. Neglecting boundary charges in force fields means ignoring electrical properties of cells.

Polarization force fields that ignore macroscopic boundary charges — no matter how sophisticated their derivation from quantum theory — cannot deal with the natural function of nerve cells as long as they depend on the coordinates of only two atoms at a time.

**Force fields of molecular dynamics: calibration**

Molecular dynamics uses force fields almost always calibrated at infinite dilution, in distilled water. That may be reasonable for the atoms inside a protein, away from mobile ions, but such a calibration must fail, in my view, for side chains of a protein that mix with mobile ions or for the mobile ions near and around proteins in bulk solution. When mobile ions are involved, screening/shielding is involved, as a general principle of physics.[8] Screening always depends on the concentration of ions, and (in nonideal cases) depends on the size and type of ions as well because 'everything' interacts with everything else in non-ideal solutions.[15, 16] Thus, force fields calibrated in distilled water will fail when dealing with concentrated solutions derived from seawater.

Seawater and the solutions of biological systems are nothing like distilled water. In fact, distilled water is lethal for nearly all cells and most proteins. Molecular dynamics computed with force fields calibrated in distilled water will have certain errors when computing proteins in physiological solutions.

**Exact equations must use the mathematics of multiscale interactions, not the mathematics of ideal solutions**

Biology occurs in modified seawater and changes in ion concentration change the reactions of most enzymes. An exact version of biochemistry must deal with ions.

I argue here that exact equations have not been possible because interactions in salt solutions that require multiscale analysis. Many types of interactions occur in ionic mixtures like seawater. All the ions in seawater are linked globally by electrostatic forces in flow, as we shall see later. Many are linked by steric interactions as well. Some are linked by orbital delocalization of electrons shared with water or other molecules, i.e., chemical bonds.

Exact theories in biochemistry must use the mathematics of interactions but that mathematics is not widely known because it has only recently been discovered.

**Interactions are not small effects**

Most biological ionic solutions, like seawater, are far too concentrated to behave like ideal fluids or electrolyte. They are, in fact, complex (not simple) fluids.[15, 16]

The free energy per mole (the experimental quantity called the activity of an ion, extensively measured in the literature [37, 39, 51, 71]) is the simplest property of an electrolyte. It is important to emphasize that activity is an experimental measurement not a theoretical construct. Physical chemists for many decades measured the activity of electrolyte solutions of a wide range of composition and concentration and showed that different methods gave similar results.[1, 10, 26, 29, 37, 52, 54, 71]

Activity plays a role something like height in a gravitational field and voltage in an electric circuit. In seawater, the activity of the bio-ions $Na^+$, $K^+$, $Cl^-$ and $Ca^{2+}$ does not vary linearly with concentration (as in an ideal fluid) or even with the square root of concentration (as in extremely dilute solutions of NaCl).[21, 38-40]

Interactions and non-ideality are not small effects in mixed ionic solutions like seawater. Interactions and non-ideality can dominate in biological systems, because ions are highly concentrated where they are most important, in and near active sites [35], ion channels, binding proteins and nucleic acids, near the 'working' electrodes of electrochemical cells, at charged boundaries in general. There, concentrations are often more than 5 molal and solution properties there are dominated by interactions.[15, 16] The activity of one ion depends on the individual concentration of every other ion. 'Everything' interacts with everything else. Some of the interactions, usually called 'allosteric' and attributed to enzymes and proteins, as structural or ensemble properties [7, 11, 28, 49], may in fact arise in the highly concentrated solutions in and near active sites of proteins.

**Mathematics of interactions**

The mathematics of interactions has been understood for a very long time when the systems involved are conservative and do not involve friction. Hamiltonians and variational calculus are the language of high-energy physicists when they build their bright X-ray sources.

Hamiltonians have not been used in most biological systems because biology occurs in condensed phases where friction is always present. Until recently, no one knew how to use Hamiltonians in systems with friction. Friction accompanies all ionic movement and conformation changes in biology because atomic collisions occur on a $10^{-16}$ sec time scale in solutions containing little empty space— that is why solutions are called 'condensed phases' — and only three or four collisions are enough to convert deterministic motion into the random motion we call heat.[6]

**Theory of complex fluids**

Recently, mathematicians have developed a theory of complex fluids that generalizes Hamiltonians into an energetic variational calculus dealing with friction. The theory has had striking successes.

Variational methods deal successfully with liquid crystals, polymeric fluids, colloids, suspensions and electrorheological fluids.[4, 5, 31, 58, 69, 70]. Variational methods describe solid balls in liquids; deformable electrolyte droplets that fission and fuse [55, 69]; and suspensions of ellipsoids, including the interfacial properties of these complex mixtures, such as surface tension and the Marangoni effects of 'oil on water' and 'tears of wine'.[25, 64, 69, 70] Variational methods allow the reformulation and understanding of problems involving interactions of considerable complexity [18, 65, 67, 68], some of which have resisted analysis for a long time. It is a little early to say the theory of complex fluids provides exact equations in general, but the theory certainly provides a productive pathway towards that goal.

The perspective offered by the variational calculus — see the tutorial presentations based on the lectures of Chun Liu [18, 24, 66] — is striking even if its results are immature. Complex fluids must be analyzed by variational methods because everything interacts with everything else. If those interactions are not addressed with mathematics, the interactions are bewildering and the results cannot be analyzed. A mathematics designed to handle interactions is needed to produce exact equations. Otherwise, interactions vary in so many ways that fixed parameters cannot deal with them [34, 37, 41, 50, 57, 71], even at infinite dilution.[30]

**Flow of charge requires global interactions and correlations**

The flow of charge at one location interacts with the flow everywhere else. Kirchoff's current law ensures correlation of charge movement everywhere, with a correlation coefficient something like 0.999 999 999 999 999 999. **The correlations produced by Kirchoff's current law are global.** Changes in flow at distant locations changes the flow everywhere.

Just consider what happens when you 'pull the plug' on an electronic device. Flow of charge into the plug ceases and atomic scale flows stop in the junctions and boundary layers of transistors. Flow on atomic scale is controlled by flows from the plug meters away. The electronic device depends on flow. The vital functions of our computers die without flow from the plug.

**Life at equilibrium is usually death**

Life also depends on flow. Flow must be dealt with consistently in biochemistry, because life does not occur without flow. Life at equilibrium is usually death. Flows of electricity are accompanied by charge imbalances that can produce large effects throughout a system. The equations of electricity are sensitive.

The equations of electricity are global, like Kirchoff's current law. The electric field in ionic solutions of living systems links everything with everything else. Exact equations must be consistent equations in which all the variables satisfy all the equations and boundary conditions in all conditions.[19, 20]

.

**An exact version of biochemistry must satisfy the equations of electricity, including global correlations of Kirchoff's current law**

The electrical forces and potentials must be computed from all charges present and their flows — in solution, in proteins and nucleic acids and macromolecules in general, in layers near lipid membranes and boundaries — because those electric forces can change qualitatively and quantitatively when net charge changes a little bit, anywhere. See the unforgettable third paragraph (p.1-1 of [23]) of "Feynman's Lectures…, Mainly Electromagnetism…") that describes how a tiny imbalance of charge is enough to lift "the entire earth."

## Continuity of Current is Exact, no matter what carries current!

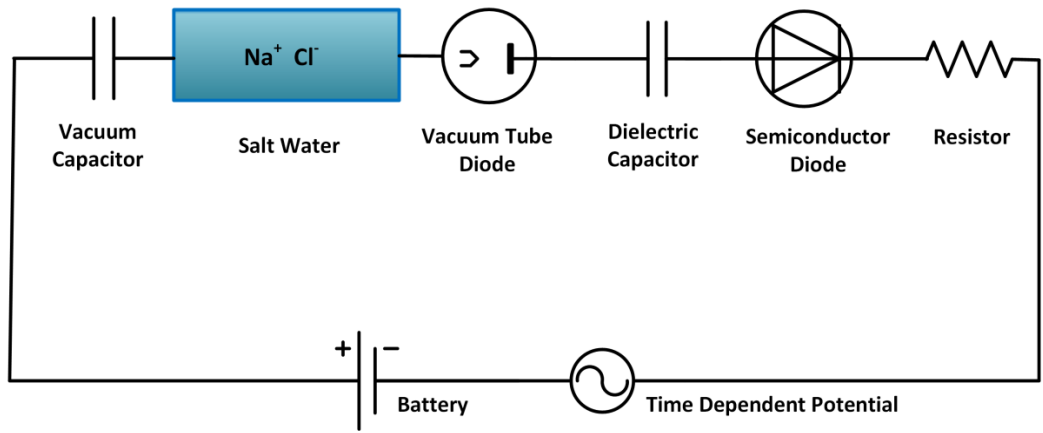

## Charge is an Abstraction, with different physics in different systems

**Charge is an abstraction: the physical nature of current is diverse**

The global dependence of the electric field is glimpsed in the cartoons of Kirchoff's current law used in computational electronics. Kirchoff's law and Maxwell's exact equations of electricity are inseparable.[2, 3, 27] Charge is the central subject of Maxwell's theory. Kirchoff's current law is really a statement of conservation of charge, including displacement charge, the abstraction and discovery of Maxwell that is his key contribution to understanding electromagnetic radiation, including light.

Charge is abstract. Charge changes physical nature as it flows through a circuit (see figure). It is ions in salt water; it is electrons in a vacuum tube; it is quasi-particles in a semiconductor; it is diverse in batteries–because of complex electrochemical reactions at electrodes; and it is nothing much in a vacuum capacitor (i.e., displacement current [27, 32]).

*The flow of current is exactly the same in every element in a series circuit, although the physical nature of that current is strikingly diverse*.

**Mass Action is about mass conservation, not charge**

Most of biochemistry describes flow by the law of mass action. The law of mass action is about mass conservation. It is not about charge conservation. The laws of electricity guarantee that the current will be the same for all reactions in series. The law of mass action does not.

The global nature of electric flow prevents the law of mass action from being exact. The law of mass action — with rate constants that are constant — does not know about charge. Its rate constants do not depend on charge in a way that guarantees Kirchoff's current law. (If you want to prove this for yourself, write the mass action equations for flux in different reactions and try to derive Kirchoff's current law as we do in the equation inset below.)

*I believe the law of mass action must be consistent with Kirchoff's current law if biochemistry is to be an exact science.*

**How do we make changes?**

How can we fix this problem and make biochemistry an exact science? How can we remake our laws so they deal well with interacting systems and electric charge? I do not know a general answer, but I know where to look for help.

Physicists for years have used consistent analysis of flow and diffusion of charges to design transistors for devices.[22, 33, 44, 47, 56, 59-63] Those devices have increased in capability by billions in the 55 years since 1959 — more precisely by $1.5^{55} = 4.8 \times 10^9$ [45, 48] — and that striking success may have something to do with the exact laws that those devices follow.

I believe biochemistry can add to its own substantial successes of the past 60 years by trying to make its laws exact. If global spatial dependence on the electric field is built into a new version of the law of mass action, along with the interactions found in nonideal solutions, we surely will do better than we have done in understanding how enzymes, channels and nucleic acids do their work.

**Consistent treatments will not be easy**

Giving up inconsistent treatments will be like giving up part of our intellectual heritage. We can no longer take the easy ways out. We can no longer look the other way when rate constants vary.

When studying allosteric interactions, we must use activities, that account for interactions among ions, not concentrations, which are appropriate only for ideal infinitely dilute solutions, because reactants and products are usually concentrated near active and allosteric sites.

We must learn to deal with fluctuating electric fields in our treatments of Brownian motion of ions so that results will not seem so anomalous.[12, 17] We can no longer compute fluctuating concentrations of charge and assume electric fields do not fluctuate.

We must incorporate boundary conditions and finite size ions into the law of mass action. Algebra and ordinary differential equations must give way to field theories, partial differential equations and variational calculus.[18, 24, 66]

We have begun that process for rate models of ionic channels [14] as I grew out of my original prejudice against such models.[9] (Stochastic analysis was responsible for that growth, more than anything else, reviewed in [14]. Applications to conduction in calcium channels are in [36, 43]. A specific example showing important consequences of long range coupling is in [42] but even there the work has just begun.)

We must even incorporate spatial inhomogeneities and electric fields into our treatments of covalent chemical reactions in ionic solution, because those spatial inhomogeneities are likely to produce very large local concentrations lasting long enough so many reactions occur at concentrations quite different from the average reactions in a spatially uniform system.

We cannot just calculate models with higher and higher resolution. We must calibrate our simulations.[46, 53] We must compute consistently with the electric field, on all scales, with theories appropriate for each scale.[13]

**Mathematics is now available**

Mathematics is finally available to deal with diffusion and electric fields in a consistent way, and the theory of complex fluids and simulations of computational electronics have shown that mathematics can describe complex fluids and devices (nearly) exactly. Now let's try that mathematics on the classical problems of biochemistry to see if we can construct a consistent theory of reactions that is exact and useful.

## Analysis

Consider the nonequilibrium situation, for a reaction where reactants and products are at different locations

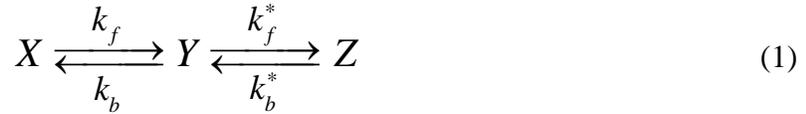

$$X \underset{k_b}{\overset{k_f}{\rightleftharpoons}} Y \underset{k_b^*}{\overset{k_f^*}{\rightleftharpoons}} Z \quad (1)$$

The flux of reactants (in a unit cross sectional area) is

$$J_{XY} = k_f [X] - k_b [Y]; \qquad J_{YZ} = k_f^* [Y] - k_b^* [Z] \quad (2)$$

where forward and backward rate constants are defined by subscripts with an asterisk for the right hand reaction, and brackets [ ] denote number density, i.e., concentration.

The flow of electric charge (in a unit cross sectional area) is current given by

$$I_{XY} = Fz_X \cdot k_f [X] - Fz_Y \cdot k_b [Y]; \quad I_{YZ} = Fz_Y \cdot k_f^* [Y] - Fz_Z \cdot k_b^* [Z] \quad (3)$$

where $F$ is Faraday's constant; $z_X, z_Y, z_Z$ are the charges on one molecule of the reactants and products. These currents are obviously not always equal even though Kirchoff's current law says they must be equal under all conditions. Algebra shows they can be equal only under special conditions:

$$I_{XY} \stackrel{?}{=} I_{YZ}$$
$$\text{if and only if}$$
$$z_X k_f [X] - z_Y k_b [Y] \stackrel{?}{=} z_Y k_f^* [Y] - z_Z k_b^* [Z] \quad (4)$$

Of course, experiments can be done under conditions that approximate the special condition of eq. (4), Then the law of mass action and Kirchoff's current law will be in approximate agreement under those conditions.

It may be possible to find a functional for rate constants which reconciles mass action with the electric field, but of course that functional must include ions throughout the system, as well as interactions of all sorts including with distant induced charge on lipid membranes and other boundaries.

We seek a global version of mass action that automatically satisfies Kirchoff's current law under all conditions.

# Acknowledgement

This is a documented, expanded version of a paper with a similar title scheduled for publication in the October 2014 issue of ASBMB Today, Editor: Angela Hopp. Chair of Editorial Advisory Board: Charlie Brenner. http://www.asbmb.org/asbmbtoday

Charlie and Angela have contributed much more to this paper than its conception and title. Fred Cohen and Tom DeCoursey made most helpful suggestions. Many thanks to them all!

I alone, of course, am responsible and for shadows cast by the particular stark light with which I view the classical landscape of chemical reactions. I am happy to fill in shadows and correct ambiguities and errors as you send them to me at my email address beisenbe@rush.edu or bob.eisenberg@gmail.com.